\documentstyle[12pt]{article}

\begin{document}

\baselineskip=15pt

\begin{flushright}
SLAC-PUB-7964\\
October 1998\\
hep-ph/9810436
\end{flushright}

\medskip

\baselineskip=18pt

\begin{center}
{\large\bf Possible Origin of Fermion Chirality and Gut Structure \\
 From Extra  Dimensions}
\footnote{Research
partially supported by the Department of Energy under
contract DE-AC03-76SF00515 }

\vspace{0.5cm}

\baselineskip=18pt

{\large Guy F. de T\'eramond}\footnote{E-mail:\quad gdt@asterix.crnet.cr}

\vspace{0.5cm} {\em \it Stanford Linear Accelerator Center,
Stanford University \\ Stanford, California 94309 \\ and \\
Escuela de F\'{\i}sica, Universidad de Costa Rica \\ San Jos\'e,
Costa Rica} \vspace{1cm}

{\bf Abstract}

\end{center}

\baselineskip=18pt

{\small \quad The fundamental chiral nature of
the observed quarks and leptons  and the emergence of the
gauge group itself are most puzzling aspects of the standard
model. Starting from general considerations of topological
properties of gauge field configurations in higher space-time
dimensions, it is shown that the existence of non-trivial
structures in ten dimensions would determine a class of models
corresponding to a grand unified GUT structure with complex
fermion representations with respect to $ SU(3)_C \otimes SU(2)_L
\otimes U(1)_Y$. The discussion is carried out within the
framework of string theories with characteristic energy scales
below the Planck mass. Avoidance of  topological
obstructions  upon continuous deformation of field
configurations leads to global chiral symmetry
breaking of the underlying fundamental theory,
imposes rigorous restrictions on the structure of the vacuum
and space-time itself and determines uniquely the gauge
structure and matter content. }

\vspace{0.1cm}

\begin{center}
(Accepted for Publication in Physical Review D)
\end{center}

\newpage

\renewcommand{\baselinestretch}{2}

{}From the perspective of theories in higher dimensions we observe zero
modes of particle fields, which are protected by the symmetries of the
theory from getting masses from a high energy scale or Kaluza-Klein
excitations.  Quarks and leptons of given helicity form a complex
representation of the gauge group, and the gauge invariance of the
theory forbids the fermions from acquiring masses at the tree
level since there is no pairing of left and right-handed fermions
with the same quantum numbers.
Although a perplexing aspect of the standard model, the
assignment of left-handed particles into doublets of the weak-isospin
group and right-handed into singlets, is crucial for the occurrence of
the fermion spectrum at low energies. A family of quarks and leptons $
u, d, e^{-}, \nu $ form a complex representation of 15 fields, not
equivalent to its complex conjugate, which transforms under $ SU(3)_C
\otimes SU(2)_L \otimes U(1)_Y$ in a highly reducible
representation $(3,2) \oplus (\bar 3,1) \oplus (\bar 3,1) \oplus (1,2)
\oplus (1,1), $ with weak hypercharges $ Y = 1/6, -2/3, 1/3, -1/2, 1 $
respectively. This is the minimal set of fields which is  free
from chiral anomalies, a condition for the renormalizability of the
theory.

In a grand unified theory~\cite{georgi-glashow}, the different
gauge interactions are embedded in a simple Lie-group whose
symmetry is manifest at a larger energy scale, and the quantum
number content of a given representation, which is arbitrary in
the standard model, follows from the transformation properties of
the unified gauge group~\cite{gutrev}. In the minimal grand
unified theory~\cite{georgi-glashow} each family of fermions is
assigned to the 15-dimensional representation  ${\bf{\bar 5+10}}$
of $SU(5)$. A family can also be assigned to an irreducible
complex representation: the spinorial ${\bf 16}$ of $SO(10)$
~\cite{georgi-fritzsh-minkowski}, or the fundamental ${\bf 27}$ of
$ E_6 $~\cite{gursey-ramond-sikivie}. Other groups which exhibit
some attractive features, have only real representations with
respect to  $ SU(3)_C \otimes SU(2)_L \otimes U(1)_Y$ and no
mechanism prevent the fermions from acquiring large masses. As an
example, the spinor representations of the orthogonal series
$SO(10+4N)$, $N > 0$, which otherwise would incorporate the
unification of generations in a simple Lie group~\cite{gutfam},
are real with respect to $ SO(10) $ and conjugate families of
unobserved fermions appear.  The same problem arises with $E_7$
and with $E_8$, the largest group of the exceptional series. Since
the fundamental chiral nature of quarks and leptons could not
depend on the details of the theory, but rather on  general or
qualitative properties of a class of theories,  they  should
belong to a universality class~\cite{witten1}. Trying to
understand the origin of chirality could thus amount to find the
global or topological properties which characterizes such
theories.

As a consequence of string dualities~\cite{polchinski} the
fundamental string scale is no longer necessarily tied to the
Planck scale~\cite{lykken,dine}, but could be anywhere between the
Planck mass and the electroweak scale according to the value of
the ten-dimensional dilaton field in the vacuum. In particular,
heterotic strings at strong couplings are equivalent to weakly
coupled Type I strings~\cite{witten2} with the possibility of
lowering the string scale to energies as low as the electroweak
scale, with large extra space-time dimensions and unique
signatures~\cite{antoniadis}. Unification of gravity and gauge
forces at the weak scale implies radical changes in the
gravitational forces at short distances, avoid the gauge hierarchy
problem altogether and even the need for low energy supersymmetry,
which has been hitherto unobserved~\cite{nima}. Large extra
spacetime dimensions have also the remarkable consequence of
changing the behavior of the running gauge coupling constants from
logarithmic to power-law above $M_o = 1/R$, where $R$ represents the
size of the extra dimensions~\cite{dienes}. As a result,
a four-dimensional GUT at scale
of $10^{16}$ GeV, the perturbative unification scale, is replaced
by a GUT in a higher dimensional space at a much lower energy
scale. This new interpretation could also give a simple
explanation of the fermion mass hierarchy~\cite{dienes}, and could
offer eventually a unified view of gauge and gravitational
unification at a TeV scale within the context of Type I
strings~\cite{shiu}. It should be noted, however, that two
basically different approaches have been followed according to
whether the extra dimensions are felt by  gauge and
gravitational interactions~\cite{antoniadis,dienes} or by the
gravitational forces only~\cite{nima}, according to different
forms of incorporating the interactions in the underlying string
theory.

In this paper we study the global
chiral symmetry properties of the ground state of a
class of theories with complex fermion representations with
respect to  $ SU(3)_C \otimes SU(2)_L \otimes U(1)_Y $. The
discussion is carried out within the framework of reduced-scale
strings where the extra dimensions are felt by all the gauge
interactions, but the results are to a large extent independent of
the particular value of the compactification scale or even the
compactification mechanism, and depend rather on general or global
properties of the underlying theory.

At energy scales well above $M_o$ and up to the Plank scale $M_P$,
where the theory is  higher dimensional and space-time appears
effectively flat, the existence of non-trivial topological
structures in ten-dimensional space-time determines a class of
models, which corresponds to  topologically stable
non-perturbative vacuum solutions that break the global chiral
symmetry of the theory. This result is related to global gauge
transformations by tunneling events between not equivalent gauge
configurations which give rise generally to topological
obstructions in the fermionic effective functional action.

The global transformations of the effective action in 10 dimensions
are expressed in terms of the 11-dimensional spectral flow of
zero modes. The restrictions imposed by the existence of topological
obstructions leads to global chiral symmetry breaking of
the underlying theory by 10-dimensional extended field
configurations, and it is shown that those restrictions
lead in fact to a unique solution.  We further study
the vacuum structure of ten-dimensional
space-time upon global chiral symmetry breaking in terms of the
particular embedding of the spin connection in the ten-dimensional
manifold, arising from the maximal subalgebra decomposition of the
original symmetry group. Poincar\'e invariance and breaking of CPT
are important issues of the model.

In the usual compactification scheme with the string scale around the
Planck scale
$M_P$, the vacuum state is assumed to be a product of $M^4 \times K$
with $M^4$ the four-dimensional space and $K$ a compact manifold with
radius of order of $1/M_P$. In the scenario with the string
scale around the electroweak scale, the physical system has strikingly
different behavior according to whether the scale $M$ is smaller or
greater than $M_o$. If $M < M_o$, the effect of  Kaluza-Klein states and additional
dimensions are ignored, whereas for scales well above $M_o$, the
effect of the Kaluza-Klein modes changes the scale-dependence of the
running couplings from logarithmic to power-law thus lowering the
scale of gauge coupling unification~\cite{dienes}.
In this limit, the size
of additional space-time dimensions appears as infinite with
respect to the scale $M$, and thus for $ M \gg M_o$ space-time
appears effectively as a 10-dimensional flat  space $M^{10}$,
greatly simplifying the description of the vacuum.  We will carry the
discussion in terms of an  effective field theory to
describe the coupling of the fermions to background fields
in the flat 10-dimensional space and ignore the
Riemannian connection.

Before proceeding we briefly review some basic relations useful
for our discussion. The Euclidean effective functional action $
exp \ (- W [A])$ describing the coupling of gauge fields to fermions
in a complex representation of a gauge group $G$ in a
2n-dimensional Euclidean manifold $M^{2n}$ is
 $$ {exp  (- W [A])}
= \int \ [d\psi] [d \overline{\psi}] \  exp \left({ - \int d^{2n}x\
 \overline{\psi}  \;/\llap D   {1\over 2} ({1 - {\overline \Gamma}} )
 \psi }\right), $$
where $\;/\llap D = \Gamma^A D_A$, the $ \Gamma_A $ are
2n-dimensional gamma matrices $({\small A,B} = 1 ... 2n)$ and
${1\over 2} ({1 - {\overline \Gamma})} $ the chirality projection
operator, with ${\overline\Gamma} = \Gamma_1 \ \Gamma_2 ...
\Gamma_{2n}$. A finite, time independent gauge transformation is
defined by the nontrivial wrapping of the map $g(x)$ around the
gauge group  $G$ for the one-form gauge connection $A = A_B \ dx^B \
$ on $M^{2n}$: $ A^g = g^{-1} A \ g + g^{-1} d \ g$, where $ d =
(\partial / \partial x^A) dx^A$, and $ g(x) \rightarrow 1$ as
$\vert x \vert \rightarrow \infty $, such that $ g(x) $ cannot be
deformed to the identity. Nontrivial field configurations are
determined by the mappings from $S^{2n -1}$ into $G$ and thus
$\Pi_{2n-1}(G)$ classifies  G-bundles over $S^{2n}$. In four
dimensions, the mappings from $S^3$ into $G$ are classified by
$\Pi_3(G)$ the third homotopy group, which is equal to the group
of integers ${\bf Z}$ for any simple Lie group. Nontrivial field
configurations in four dimensions (instantons) are homotopically
equivalent since according to a theorem due to Bott~\cite{bott},
any continuous mapping of $S^3$ into a simple Lie group can be
continuously deformed in a mapping into an $SU(2)$ subgroup of
$G$, equivalent to an $S^3 \rightarrow S^3$ mapping, and thus it
is not possible to differentiate among topologically nontrivial
configurations from different groups. On the other hand, theories
in higher dimensions present a diversity of topological structure
which is, from the point of view of the homotopy properties,
absent in a four dimensional space.

The coupling of fermion zero modes to the background field
strength \mbox{$F=dA + A \land A$} $ = {1\over 2} F_{AB} \ dx^A \land
dx^B$, is expressed in terms of the character-valued index of the
Dirac operator~\cite{index} \ which is the difference of fermion
zero modes of opposite chirality: \mbox{$\psi^+ =  {1\over 2} ({1 +
{\overline \Gamma}}) \psi$} and  $\psi^- =  {1\over 2} ({1 -
{\overline \Gamma}}) \psi$:
$$ ind \;/\llap D_{2n} = n^+ - n^- = \int \ ch(F) = \int \ {1\over
n !} \ {({i \over {2 \pi}})^n} \  Tr \ F^n ,$$
where the integral is over a 2n-dimensional manifold and the trace
is taken for states in a given fermion representation of $G$. The
Chern character is given by $ ch(F) = Tr \ e^{iF/2 \pi} $. Since $
Tr \ F^n$ is a closed 2n-form,  $d Tr \ F^n$ is locally exact and
can be expressed as a $2n-1$ Chern-Simons form: $Tr \ F^n = d
\omega_{2n-1}(A)$, with $ \omega_{2n-1}(A) = n \ \int^1_0 \  \
Tr[A \ F^{n-1}_t] \ dt$ and $F_t = t \ dA + t^2 A^2$. Using
Stoke's theorem, and integrating by parts~\cite{index}
$$ ind \;/\llap D_{2n} = \ \int_{S^{2n-1}} Q_{2n-1} ( g^{-1} d g)
= {(n - 1)!\over{2^n \pi^n (2n - 1)!}} \ \int_{S^{2n-1}}  { \ Tr (
g^{-1} d g)^{2n-1}} ,   $$
where $Q_{2n-1} = {1/ n !} \ {(i / 2 \pi)}^n \ \omega_{2n-1}$. The
right hand side of the above expression is an integer representing
the winding number or topological charge of the homotopy classes
of $\Pi_{2n-1}(G)$, which depends only on the properties of
$g(x)$.

In 2p-dimensions, the Chern class is given in terms of the 2p-form
$ d_{a_1 ...a_p}\ F^{a_1} \land ... \land F^{a_p}$, where $ d_{a_1
...a_p}$ is a totally symmetric invariant tensor. In terms of the
representation matrices $X$ of an irreducible representation of
the generators of the Lie algebra of $G$, $d$ is written as a
symmetric trace denoted by $STr$, which is a trace over all the
permutations of the product of $p$ representation matrices: $ d_{a_1
...a_p} = STr(X_{a_1} ... X_{a_p})$. The operator
$$ I_p = {1\over{p!}} \sum_{a_1 ...a_p} \ d^{a_1 ...a_p} \ X_{a_1}
... X_{a_p}, $$
commutes  with all the elements of the algebra in a given
representation, and is in fact a Casimir invariant of order $p$. If
the Lie algebra has no Casimir invariant of order $p$ $ ( d_{a_1
...a_p} =0)$, the index theorem implies that there are no fermion
zero modes of definite chirality.

Extended field configurations which carry a conserved quantum
number are stable and the possible field configurations are
determined by the condition that some functional of the fields is
finite~\cite{weinbergbook}. Such configurations are characterized
by their global properties in terms of the elements of homotopy
groups. To understand better the relation between a Casimir
invariant and homotopy, as well as disposing of a simple tool for
finding the properties of the homotopy groups, we express a
compact connected Lie group as a product of spheres following
Pontrjagin and Hopf. This useful expansion, allows us to determine
the topological properties of simple Lie groups by reading the
properties of the mappings between spheres~\cite{boya}.
The rank of $G$ is equal to the number of Casimir
invariants and the number of spheres in the expansion is equal to the
rank of the group.  A Lie
group $G$ of rank $m$ behaves as the product of m odd-dimensional
spheres $S^{2p-1}$, with $p$ the order of each Casimir invariant in
$G$.  As an example the rank 2 group $SU(3)$ has
Casimir invariants of order 2 and 3 and the exceptional rank 4
group $F_4$ of order 2, 6, 8 and 12. Thus, $SU(3)$ and $F_4$ are
expressed as $SU(3) \sim S^3 \times S^5$ with dimension 8 and $F_4
\sim S^3 \times S^{11} \times S^{15}\times S^{23}$ with dimension
52. Since the mappings from $S^{n} \to S^{n}$ are classified
by $\Pi_n(S^n) = \bf Z$, it
follows that the existence of a Casimir of order $p$ in the algebra
of a compact group $G$ is equivalent to the condition $\Pi_{2p-1}(G)
=  \bf Z$.

Consider a theory in a 2n-dimensional Euclidean manifold $M^{2n}$
and the coupling of gauge fields to fermions in a  representation
of a gauge group $J$ (not necessarily complex)  described by the
Euclidean effective functional action $exp( - W [A^J])$, such that
the theory has trivial topology in $2n$ dimensions: $ \Pi_{2n-1}(J)
= 0$. We further assume that the underlying theory is free from
perturbative anomalies, gauge and gravitational, before
compactification or symmetry breaking. According to the discussion
above, the operator $\;/\llap D(A^J)$ has no zero modes of
definite chirality and furthermore we can set \mbox {$A^J = 0$} in the
vacuum. However the existence of nontrivial topological
structures in  higher
dimensions~\cite{horvath}, could imply that the vacuum state with
all the gauge fields vanishing simultaneously is not a stable
solution. If the theory has nontrivial field configurations for a
subgroup $G$ of $J$, $G \subset J$, $\Pi_{2n-1}(G) \not = 0$, the
symmetry can be broken in the sector of the theory characterized
by the non-trivial G-bundle. Equivalently, it can be stated that
the theory is broken at the quantum level from the fermionic
measure $[d\psi] [d \overline{\psi]}$ in the effective Euclidean
functional action.

To obtain the functional integral $exp(- W [A^G])$  from $exp(- W
[A^J])$ we must consider the transition  from $A^J$ to $A^G$,
which is equivalent to deform continuously one field configuration
into the other, while keeping the effective action $W[A]$ finite.
Since we are considering disconnected gauge transformations, the
interpolation of configurations cannot be done along the
connection fiber, but rather we should consider a one-parameter
family $A_t$ interpolating from $A^J$ to $A^G$ along a path
defined in the space of gauge connections ${\cal C}$, which is
contractible. We can take for example $A_t = A^J + t(A^G - A^J)$
with $ 0 \leq t \leq 1$, which corresponds to an instanton
transition in the temporal gauge. Although ${\cal C}$ is
contractible, topological obstructions will generally appear in
the functional action  which must be integrated over the gauge orbit
space, ${\cal C/G}$, the space of all gauge not equivalent
configurations  (${ \cal G} $ is the group of all gauge
transformations). The space ${\cal C/G} $ is in general
non-contractible and gives rise to the non-trivial topology of the
theory as we interpolate the effective action from $t=0$ to $t=1$ as a
functional of $A_t$:
$$\Delta W = \int^1_0 \ dt \ {d\over{dt}}  W [A_t] = W [A^G] -  W
[A^J].$$

The  change in the effective action $\Delta W$ is obtained by
integrating out the fermionic fields and is given by a general
formula derived by Witten in his study of global
anomalies~\cite{global}
$$   \Delta W = {{\pi i} \over 2 }\eta \ \ (mod  \ 2 \pi i) ,$$
where $\eta$ is the invariant of Atiyah, Patodi and
Singer~\cite{aps}, which expresses the spectral asymmetry of the
eigenvalues $\lambda$ of the Hamiltonian
$$\eta = {\lim_{\epsilon \to 0}} \sum_{\lambda} \ sgn(\lambda) \
exp( - \epsilon \vert \lambda \vert),$$
on a $2n +1$ dimensional manifold $M^{2n} \times S^1$.

It is very useful to represent the spectral asymmetry in $2n+1$
dimensions as the continuous change in $\eta$, and express the
spectral flow in terms of an index in $2n+2$ dimensions
~\cite{aps,gaume-dellapietra-moore}. In terms of the gauge field $A$
defined in the $2n$ manifold $M^{2n}$ we can define a field in a
$2n+1$ dimensional manifold $M^{2n} \times S^1$ as ${\cal A} =
(A,0)$, as well as an interpolating $2n+1$ field ${\cal A}_t$ by
${\cal A}_t = {\cal A}^J + t ({\cal A}^G - {\cal A}^J)$, $ 0 \leq
t \leq 1$. Using the results of
Ref.~\cite{gaume-dellapietra-moore}, we write the change in the
effective functional action  in terms of the difference between
$2n+1$ Chern-Simons at the end-points of the path of integration
along the variable $t$:
$$exp{\ \left\{-( W[A^G] \ - \ W[A^J])\right\} } = exp \ \left[{{ - {i \over 2}
\pi } {\int_{M^{2n} \times S^1}  Q_{2n+1} ({\cal A}^G ) \ - \
Q_{2n+1} ({\cal A}^J)}}\right]. $$

Topological obstructions arise unless $Q_{2n+1}({\cal A}^G )$ and
$Q_{2n+1} ({\cal A}^J)$ vanish. In terms of homotopy this
condition is equivalent to $\Pi_{2n+1}(G) = 0$ and $\Pi_{2n+1}(J)
= 0$, since the $2n+1$ dimensional spectral flow and the $2n+2$
dimensional index are represented by the same invariant in the
absence of gravity~\cite{gaume-dellapietra-moore}.

If the effective field theory at scales $M \gg M_o$ is embedded in
a weak coupling Type I String, we could take advantage of the
duality maps between strings to examine the strong coupling regime
of the equivalent heterotic string with gauge group $E_8 \otimes
E_8$ ~\cite{ghmr}. Since the topology of $E_8$ in ten dimensions
is trivial, $\Pi_9(E_8) = 0$, the gauge fields associated with one
of the $E_8$ groups in the vacuum could be set equal to zero and
decouple from the fermion sector. As discussed above, the
existence of non-trivial bundles in a sector of $E_8$ could imply
that the non-perturbative vacuum structure cannot be a state with
all the gauge fields vanishing simultaneously. We assume that
the trivial vacuum is indeed broken by non-trivial topological
structures given in terms of some finite functional of the fields.
The possible field configurations are classified by their homotopy
properties and are determined by the ninth homotopy group of the
gauge group $G$, $\Pi_9(G)$, in ten dimensions.

In a survey of global properties of the compact connected
simple Lie groups~\cite{dict} it is found that:
$$ \Pi_9(SU(5)) = {\bf Z}, \quad  \Pi_9(SO(10)) = {\bf Z} + {\bf
Z}_2, \quad \Pi_9(E_6) = {\bf Z}, $$
which correspond precisely to the GUT theories with complex
fermion representations with respect to $ SU(3)_C \otimes SU(2)_L
\otimes U(1)_Y$~\cite{teramond}. The higher dimensional groups $
SU(N), \quad N \ge 6$, which also have complex fermions
representations with respect to SU(5) belong to the same class.
$\Pi_9$ is stabilized for the unitary groups at $N=5$, which
in fact is the lowest rank group that incorporates the strong and electromagnetic
interactions. The groups $SO(10)$ and $E_6$, which are successful
embedding the different integration subgroups into a larger rank
group, belong to the same class. For all other Lie groups $\Pi_9$
has a finite number of elements or it is trivial~\cite{dict}. In
terms of representation theory, as discussed above, the condition
$ \Pi_9(G) = {\bf Z} $ is equivalent to find out among all compact
Lie groups, which have a fifth-order Casimir
invariant~\cite{tosa}. Note also the existence of fermion zero
modes with definite chirality in ten dimensions for the class of
GUT models with complex fermion representations, as consequence of
the index theorem in the presence of non-trivial field
configurations.

Let us finally examine the topological obstructions in the
effective functional action from the global chiral symmetry
breaking of the theory in ten dimensions, as we interpolate
between states by continuously deforming the field configurations
in the functional integral $exp( - W[A])$. As discussed above,
the change in the effective functional action is expressed
in terms of the 11-dimensional spectral flow of zero modes.
The theory has no topological obstructions if the Chern-Simons form
$Q_{11}({\cal A}_t)$ vanishes for the initial and final
configurations. Equivalently, it can be stated that the theory has
no topological obstructions under global transformations between
gauge non equivalent configurations  if $\Pi_{11}$ is zero for the
initial and final gauge groups. Examining the global properties
of the compact connected Lie groups we find the remarkable
result that among an infinite number of possibilities
$\Pi_{11}$ is zero only for $E_8$ and $SU(5)$~\cite{dict}:
$$ \Pi_{11}(E_8) = 0 \quad and \quad \Pi_{11}(SU(5)) = 0, $$
allowing a transition from $E_8$, the gauge group of the
heterotic string, to $SU(5)$, the minimal grand unified theory
that contains the observed quarks and leptons.

According to the CPT theorem  the fermions of given
chirality in ten dimensions transform under real representations
of the gauge group. Supersymmetry requires that this
representation is the adjoint. Both conditions are met by $E_8$,
since its fundamental representation is its adjoint. The transition
from $E_8$ to $SU(5)$, a group with complex fermion representations,
breaks the chirality of the theory and furthermore supersymmetry
is not preserved by the transformation. Note that CPT is also broken.
Since $E_8$ has real fermionic
representations, according to the character-valued index theorem
of the Dirac operator, the appearance of fermion zero modes of
definite chirality in ten dimensions signals the breaking of the
global chiral symmetry of the theory.

Under the maximal subalgebra decomposition, $E_8$ breaks as $SU(3)
\otimes E_6$, $SO(16) \supset  SO(6) \otimes SO(10)$ and $SU(5)
\otimes SU(5)$. In the usual description of string
compactification of higher dimensions~\cite{stringbook}, the
vacuum is assumed to be a product of the form  $ M^4 \times K $ to
preserve four-dimensional Poincar\'e invariance and a vacuum
expectation value is introduced on $K$ to break $E_8$, otherwise
there is no chiral asymmetry of fermions in four dimensions. This
is accomplished by identifying the spin connection of $K$ with
background gauge fields on $SO(6)$ or $SU(3)$, which act as
holonomy group of the manifold $K$, the group of rotations in the
tangent space generated by parallel transport on closed loops on
$K$.

In the present work, the $E_8$ symmetry is not broken by
compactification but rather as consequence of the breaking of
global chiral symmetry. As discussed above, this is only possible
for a symmetry breaking along $SU(5) \otimes SU(5)$  since
$$\Pi_{11}(E_8) = 0 \quad and \quad \Pi_{11}(SU(5) \otimes SU(5)) =
\Pi_{11}(SU(5)) \otimes \Pi_{11}(SU(5)) = 0.$$

Note that both $SU(5)$ groups have non-trivial configurations in
ten dimensions. One SU(5) carries the GUT structure and its
algebra the quantum numbers of the fermion representations, the
other $SU(5)$ constitutes a natural embedding in a 10-dimensional
manifold, where the spin connection is an $O(10)$ gauge field and
the holonomy group is a subgroup of $O(10)$. If the spin
connection $w$ is identified with a background gauge field on the
second $SU(5)$, acting as holonomy group, the configurations for
the two-form field $F$ and the Lie-algebra valued curvature
two-form $\Omega = dw + w \land w$ are determined by the integer
classes, $C_n = \int  c_n$ , corresponding to a fifth Chern class
$c_5$ in a ten dimensional manifold:
$$ {i\over{3840}}\ \int_{M^{10}} \ F \land F \land F \land F \land
F = C_5,$$
$$ {i\over{3840}}\ \int_{M^{10}} \ \Omega \land \Omega \land
\Omega \land \Omega \land \Omega = C_5.$$

To describe the particular embedding of the second SU(5), it is
convenient to introduce a complex manifold $K^n$, of dimension n,
with holomorphic transition functions in a real manifold $M^{2n}$,
of dimension $2n$~\cite{eguchi}. As an example, Euclidean
2n-dimensional space is identical to the n-dimensional complex
manifold $C^n$ with a complex metric. In a K\"ahler manifold, the
vector representation of $SO(2N)$ is given in terms of the
fundamental representations $ N \oplus {\overline N}$ of the
holonomy group $U(N)$, according to the real K\"ahler two-form
harmonic metric $K = {i \over 2} g_{a \overline b} d z ^a \land d
{\overline z}^{\overline b}$, where
$$  \int_{K^n} \ K \land K \land ... \land K \ > 0,$$
defines the natural orientation of the manifold. If a K\"ahler
manifold  $K^n$ has a metric of $SU(N)$ holonomy, the first Chern
class of the manifold $c_1$ is zero and the metric is Calabi-Yau.
Our particular embedding has a Calabi-Yau metric of SU(5)
holonomy, and since a Calabi-Yau manifold is Ricci-flat it
corresponds to a 10-dimensional manifold with cosmological
constant zero. The K\"ahler form $K$ can be expressed locally in
terms of a zero form $\phi$, the K\"ahler potential, as $ K = {i
\over 2} \ \partial {\overline \partial} \phi$.  $K$ is in fact the
curvature of the manifold: $ \Omega = 2i K$.

To obtain a physical theory in four dimensions Poincar\'e
invariance should be recovered at large distances away from the
extended field configurations or 10-dimensional instantons. In
terms of the variables $z_1 = x + iy$, $z_2 = z + it$, $z_3$,
$z_4$ and $z_5$ this implies that $\phi$ is flat along the $z_1$
and $z_2$ direction
$$ \phi(z_1,...,z_5)_{z_1,z_2 \to \infty} \to \ \vert z_1 \vert^2
+ \vert z_2 \vert^2,$$
and bounded by a distance R along the other complex coordinates.
This configuration of space-time with extra dimensions is
reminiscent of the domain wall interpretation  given some time ago
by Rubakov and Shaposhnikov~\cite{rubakov} of the kink solution in
a five dimensional space, leading to fermion zero modes in four
dimensions. The actual solution in ten dimensions would follow
from the 10-form equation given above in terms of the curvature
$\Omega$, with $c_1(\Omega) = 0$. The solution of the $SU(5)$
instanton $F$ should be studied along similar lines, but without a
restriction on the first Chern class, $c_1(F)$. Hopefully, only
the diagonal group generators survive away from the gauge
instanton, thus breaking $SU(5)$ and avoiding the proton decay
problem. A thorough study of the ten-dimensional extended field
configurations or instantons, including the stability of the
solutions and the relation between the metric and field
configurations, should be undertaken to confirm the theoretical
viability of the ideas here exposed. In particular, the
determination of the spectrum of fermion zero modes
in four dimensional space-time is
crucial. Experimentally, a signature of the global chiral symmetry
breaking mechanism discussed in this paper include effects from
the breaking of CPT and Poincar\'e invariance which are suppressed
by a factor of order of $M_o/M_P$~\cite{kostelecky}.

A major reason for the introduction of grand unified models as
well as the motivation behind contemporary studies of Kaluza-Klein
theories is the need to understand the quantum numbers of  quarks and
leptons and their fundamental chiral nature. We have shown in this
paper how both approaches are related. We have
used topological considerations as valuable tools for studying
general properties of physics in higher dimensions due to the
richness of nontrivial structures that are present. The existence of
nontrivial topological structures give us an indication of
the origin of the structure of grand unified theories and the
observed spectrum of chiral fermions, which is protected from
acquiring Kaluza-Klein excitations by the coupling of zero modes
to nontrivial background field configurations. Breaking of the global
chiral symmetry of the theory follows from the continuous deformation
of field configurations in the fermionic functional action,
leading to a unique solution imposed by the avoidance of
topological obstructions. The solution found could also have profound
implications in the structure
of space-time itself as supported by extended field configurations
with non trivial topology in the metric and the gauge fields
and could constitute a viable alternative to
compactification mechanisms.

I have benefited from discussions with E. Silverstein, M. E. Peskin,
N. Arkani-Hamed, R. Espinoza,  J.M. Rodr\'{\i}guez and J. Varilly.

\vfill
\eject


\begin{thebibliography}{100}

\bibitem{georgi-glashow}
H. Georgi and S. L. Glashow, Phys. Rev. Lett. {\bf32}, 438 (1974).

\bibitem{gutrev}
For a review, see P. Langacker, Phys. Rep. {\bf 72}, 185 (1981);
R. Slansky, Phys. Rep. {\bf 79}, 1 (1981);
X. C. de la Ossa and G. F. de T\'eramond, Ann. Phys. (N.Y.), {\bf 155}, 358
(1984).


\bibitem{georgi-fritzsh-minkowski}
H. Georgi, in {\it Particles and Fields - 1974}, proceedings of
the 1974 meeting of the APS Division of Particles and Fields, ed. C. Carlson, AIP
Conf. Proc. No 23 (AIP, New York, 1975); H. Fritzsh and P. Minkowski,
Ann. Phys. (N.Y.) {\bf 93}, 193 (1975).

\bibitem{gursey-ramond-sikivie}
F. G\"ursey, P. Ramond, and P. Sikivie, Phys. Lett. {\bf 60B}, 177 (1976);
F. G\"ursey and P. Sikivie, Phys. Rev. Lett. {\bf 36}, 775 (1976).

\bibitem{gutfam}
M. Gell-Mann, P. Ramond, and R. Slansky, in
{\it Supergravity}, eds. P. Van Nieuwenhuizen and D.Z. Freedman
(North-Holland, Amsterdam, 1979); F. Wilczek, in {\it Proceedings of the 1979
International Symposium on Lepton and Photon Interactions at High
Energies}, ed. by T. Kirk and H. Abarbanel (Fermilab, Illinois, 1980).

\bibitem{witten1}
E. Witten, {\it Proceedings of the Workshop on Unified String
Theories} (Santa Barbara, California, 1985).

\bibitem{polchinski}
For a review, see J. Polchinski, Rev. Mod. Phys. {\bf 68},  1245 (1996).

\bibitem{lykken}
J. D. Lykken, Phys. Rev. D {\bf 54}, 3693 (1996).

\bibitem{dine}
M. Dine, Y. Shirman, Phys. Lett. {\bf B377}, 36, (1996).

\bibitem{witten2}
E. Witten, Nucl. Phys. {\bf B471}, 135 (1996);
P. Horava and E. Witten, Nucl. Phys. {\bf B475}, 94 (1996).

\bibitem{antoniadis}
I. Antoniadis, Phys. Lett. {\bf B246}, 377 (1990);
I. Antoniadis, C. Mu\~noz, and M. Quir\'os, Nucl. Phys. {\bf B397}, 515
(1993); I. Antoniadis, K. Benakli, and M. Quir\'os, Phys. Lett.
{\bf B331}, 313 (1994).

\bibitem{nima}
N. Arkani-Hamed, S. Dimopoulos, and G. Dvali, hep-ph/9803315;
I. Antoniadis, N. Arkani-Hamed, S. Dimopoulos, and G. Dvali, hep-ph/9804398.

\bibitem{dienes}
K.R. Dienes, E. Dudas, and T. Gherghetta, hep-ph/9803466,
hep-ph/9806292, hep-ph/9807522.

\bibitem{shiu}
G. Shiu and S.-H.H. Tye,  hep-th/9805157. See also: C. Bachas,  hep-ph/9807415.

\bibitem{bott}
R. Bott, Bull. Soc. Math, France, {\bf84}, 251 (1956).

\bibitem{index}
M. F. Atiyah and I. M. Singer, Ann. Math. {\bf87}, 484
(1968); {\bf87}, 546 (1968); {\bf93}, 119 (1968); {\bf93}, 139 (1968);
M. F. Atiyah and G. B. Segal, Ann. Math. {\bf87}, 531 (1968).
For a review of index theorems, see B. Zumino,
in {\it Relativity, Groups and Topology II}, proceedings of the
Les Houches Summer
School, 1983, edited by B. S. de Witt and R. Stora (North Holland,
Amsterdam, 1984). See also: L. Alvarez-Gaum\'e, Lectures given
at Int. School on Mathematical
Physics, Erice, Italy, Jul. 1985 (Erice School Math. Phys, 1985);
L. Alvarez-Gaum\'e and P. Ginsparg, Ann. Phys. {\bf 161}, 423 (1985).

\bibitem{weinbergbook}
S. Weinberg, {\it The Quantum Theory of Fields II}, (Cambridge
University Press, 1996): Chapter 23.

\bibitem{boya}
For a review and explicit construcion, see L. J. Boya, Rep. on
Mathematical  Physics,  {\bf30}, 149 (1991).

\bibitem{horvath}
A. Polyakov, Nucl. Phys. {\bf B120}, 429 (1977);
Z. Horvath and L. Palla, Nucl. Phys. {\bf B142}, 327 (1978).

\bibitem{global}
E. Witten, Commun. Math Phys, {\bf 100}, 197 (1985).

\bibitem{aps}
M. F. Atiyah, V. K. Patodi and I. M. Singer,
Proc. Camb. Philos. Soc. {\bf 77}, 43 (1975); {\bf 78}, 405 (1975);
{\bf 79}, 71 (1976).

\bibitem{gaume-dellapietra-moore}
L. Alvarez-Gaum\'e, S. Della Pietra, and G. Moore,
Ann. Phys. (N.Y.), {\bf 163}, 288 (1985).

\bibitem{ghmr}
D. J. Gross, J. A.  Harvey, E.  Martinec, and R. Rohm, Nucl. Phys.
{\bf B256}, 253 (1985).

\bibitem{dict}
{\it Encyclopedic Dictionary of Mathematics}, edited by
S. Iyanaga and Y. Kawada (MIT, Cambridge, 1977), Vol. 2, 1417.

\bibitem{teramond}
G.F. de T\'eramond, Phys. Rev. D {\bf 31}, 1516 (1985).

\bibitem{tosa}
Y. Tosa and S. Okubo, Phys. Rev. D {\bf 23}, 3058 (1981).


\bibitem{stringbook}
M. B. Green, J. H. Schwarz, and E. Witten, {\it
Superstring Theory} (Cambridge University Press, 1987), Vol. 2.


\bibitem{eguchi}
T. Eguchi, P.B. Gilkey, and A. J. Hanson, Phys. Rep. {\bf 66},
213 (1980).

\bibitem{rubakov}
V. A. Rubakov and M. E. Shaposhnikov, Phys. Lett. {\bf B125}, 136,
(1983). See also: C. G. Callan and J. A. Harvey, Nucl. Phys. {\bf
250}, 427 (1985); G. Dvali and M. Shifman, Phys. Lett. {\bf
B396}, 64 (1997).

\bibitem{kostelecky}
For a review of the status of CPT and Lorentz symmetry
see: A. Kosteleck\'y, hep-ph/9810365 and references therein.

\end{thebibliography}
\end{document}